\documentclass[pra,twocolumn,showpacs,amsmath,amssymb,superscriptaddress]{revtex4-1}
 \usepackage{amsmath}
 \usepackage{graphicx}
 \usepackage{bm}
 \usepackage{bbm}
 \usepackage{subfigure}
 \usepackage{amssymb}
 \usepackage{xcolor}
 \usepackage{xcolor,xspace,colortbl,rotating}
 \usepackage{amsfonts}
 \usepackage{tabulary}
\begin{document}
	\bibliographystyle{pss}
\bibliographystyle{apsrev4-1}
\title{Concepts of antiferromagnetic spintronics}
\author{O. Gomonay}
\affiliation{Institut f\"ur Physik, Johannes Gutenberg Universit\"at Mainz, D-55099 Mainz, Germany}
\affiliation{National Technical University of Ukraine ``KPI'', 03056, Kyiv,
	Ukraine}
\author{T. Jungwirth}
\affiliation{Institute of Physics ASCR, v.v.i., Cukrovarnicka 10, 162 53 Praha 6 Czech Republic}
\affiliation{School of Physics and Astronomy, University of Nottingham, Nottingham NG7 2RD, United Kingdom}
\author{J. Sinova}
\affiliation{Institut f\"ur Physik, Johannes Gutenberg Universit\"at Mainz, D-55099 Mainz, Germany}
\affiliation{Institute of Physics ASCR, v.v.i., Cukrovarnicka 10, 162 53 Praha 6 Czech Republic}
\begin{abstract}
Antiferromagnetic spintronics is an emerging research field whose focus is on the electrical and optical control of the antiferromagnetic order parameter and its utility in information technology devices. An  example of recently discovered new concepts is the N\'{e}el spin-orbit torque which allows for the antiferromagnetic order parameter to be controlled by an electrical current in common microelectronic circuits. In this review we discuss the utility of antiferromagnets as active and supporting materials for spintronics, the interplay of antiferromagnetic spintronics with other modern research fields in condensed matter physics, and its utility in future "More than Moore" information technologies.
\end{abstract}
\maketitle

\section{Introduction: Spintronics and magnetic recording}
\label{intro}
Conventional spintronic devices are based on the control of  magnetic moments in ferromagnets \cite{Chappert2007}.  Antiferromagnetic spintronics, on the other hand, is a  field dealing with  devices whose key or important parts are made of antiferromagnets \cite{MacDonald2011,Gomonay2014,Jungwirth2016,Baltz2016}. A recent discovery of the electrical switching of an antiferromagnet by  the N\'eel spin-orbit torque \cite{Wadley2016} gives an example that antiferromagnetic moments can be controlled in microelectronic devices by means comparably efficient to ferromagnets. This opens the possibility to unlock a multitude of known and newly identified unique features of antiferromagnets for spintronics research and applications. Here we give a brief overview of this active field, based on a presentation at the Antiferromagnetic Spintronics Workshop organized by the Spin Phenomena Interdisciplinary Center in Mainz in September 2016 \cite{tomas}. 
More details can be found in several recent extensive reviews \cite{MacDonald2011,Gomonay2014, Jungwirth2016, Baltz2016}. 

The field of spintronics is tightly related to magnetic recording, the history of which started from the recording of sound. One of the first techniques was the magnetic wire recorder invented at the end of the 19th century in parallel to the gramophone. The electromagnetic coil was used for the recording and the electric field inductively generated in the same coil was  used for retrieving the recording.  Magnetic wire recorders were, however, unsuccessful in competing with the gramophones. The situation changed in 1930's when the tape recorder was invented. Using a much more practical, magnetically coated plastic tape,  it became a major technology for recording sound, and later also video and data. A further revolution in magnetic recording technologies dates to the 1950's when hard disk drives came into play. At the same time magnetic core memory was invented marking the dawn of solid-state computer memories. Its present integrated-circuit counterparts are the magnetic random access memories (MRAMs) \cite{Chappert2007} which offer the non-volatile alternative to semiconductor computer memories. Remarkably, the currently commercially available toggle MRAMs still use the same 19th century physical principle for writing data by the electromagnet.

While other recording technologies like gramophone and CDs have or are becoming obsolete, magnetic recording is a keeper.  Hard drives and magnetic tapes provide the virtually unlimited data storage space on the internet. MRAMs are among the leading candidate technologies for the "More than Moore" era that we have now formally entered after the official end in 2016 of the Moore's Law driven International Technology Roadmap for Semiconductors \cite{Waldrop2016}. However,  the 19th century inductive coils would not allow to keep magnetic recording competitive with semiconductor storage and memory devices. In hard drives, the coils were removed from the readout and replaced by spin-based magneto-resistive elements, namely by the anisotropic magnetoresistance (AMR)  and later by the giant or tunneling magnetoresistance (GMR or TMR) devices. The development of the integrated-circuit MRAMs would have been unthinkable without these spintronic readout schemes \cite{Daughton1992,Chappert2007}. 

21st century brought yet another revolution in ferromagnetic recording by eliminating the electromagnetic induction from the writing process in MRAMs and replacing it with the spin-torque phenomenon \cite{Chappert2007}. In the non-relativistic version of the effect, switching of the recording ferromagnet is achieved by electrically transferring spins from a fixed reference permanent magnet \cite{Ralph2008}. In the recently discovered relativistic version of the spin torque, the reference magnet is eliminated and the switching is triggered by the internal transfer from the linear momentum to the spin angular momentum under the applied writing current \cite{Sinova2015}. The complete absence of electromagnets or reference permanent magnets in this most advanced physical scheme for writing in ferromagnetic spintronics has served as the key for introducing the new physical concept for the efficient control of antiferromagnetic moments \cite{Zelezny2014}. 

Antiferromagnets, ferrimagents, or helical magnets have zero or small magnetization and often an intriguingly complex magnetic structure. These materials are more common than ferromagnets and can be superconductors, metals, semimetals, semiconductors, or insulators, in comparison to the primarily metallic ferromagnets. Another appealing property of antiferromagnets is the orders of magnitude faster spin-dynamics time scale than in ferromagnets. The antiferromagnetic resonance is in THz, driven by the strong exchange interaction between the spin sublattices, while the GHz ferromagnetic resonance is governed by the weak anisotropy energy. 

The simplest, two-spin-sublattice collinear antiferromagnets have the magnetic order comprising alternating (staggered) magnetic moments that compensate each others. For this reason they do not produce stray fringing fields and are weakly coupled to the external magnetic field.  For nearly a century since their discovery, antiferromagnets have been considered as magnetically inactive materials with no practical utility in devices. The recently discovered antiferromagnetic version of the relativistic spin torque \cite{Zelezny2014,Wadley2016} is among the recent achievements that have opened a way to efficiently control the aniferromagnetic order and to start the research and development of memory and other spintronic devices that can exploit the unique properties of antiferromagnets.

\section{Spin-orbitronics and N\'eel spin-orbit torque switching}
\label{NSOT}
Relativistic spin-orbit torques are attributed to spin-orbitronics phenomena that were originally discovered in non-magnetic systems, namely to the spin Hall effect (SHE) and the inverse spin galvanic (Edelstein) effect (iSGE) \cite{Sinova2015}. Both effects originate from the coupling between the spin angular momentum  and the linear momentum of an electron. In the SHE, the spin-orbit coupling induces a separation of spin-up and spin-down electrons in the direction transverse to the direction of the charge current.  In other words, due to the SHE, the electrical current creates a pure spin current in the bulk of the sample. This leads to an accumulation of opposite spins  at opposite edges of the sample which is allowed by symmetry since the edges break the invariance under space-inversion.  In the iSGE, the charge current induces directly a non-equilibrium spin polarization which is uniform within the bulk of the sample. In this case, symmetry requires that the crystal unit cell lacks the inversion center.

The SHE and iSGE were independently observed in 2004 by several groups \cite{Kato2004d,Ganichev2004a,Silov2004,Wunderlich2004,Wunderlich2005}. Both phenomena can be present simultaneously in the same structure, as was shown, e.g., in Refs.~\cite{Wunderlich2004,Wunderlich2005}. In these experiments the current-induced spin polarization was detected by  optical means. In the strongly spin-orbit coupled valence band of p-type GaAs, the measured spin-polarizations produced by  $\sim100$~mA  currents  reached $\sim1-10$~\%. Generating the same spin polarizations in this nominally non-magnetic semiconductor by conventional means required external magnetic fields of several Tesla.  The comparison between efficiencies of Dirac and Maxwell equation physics is striking here: The quantum-relativistic SHE or iSGE phenomena allow to achieve the same spin polarization with microchip currents as with a classical, $\sim 100$~A laboratory-size electromagnet. 

\begin{figure}[h]
	\centering
	\includegraphics[width=1.00\columnwidth]{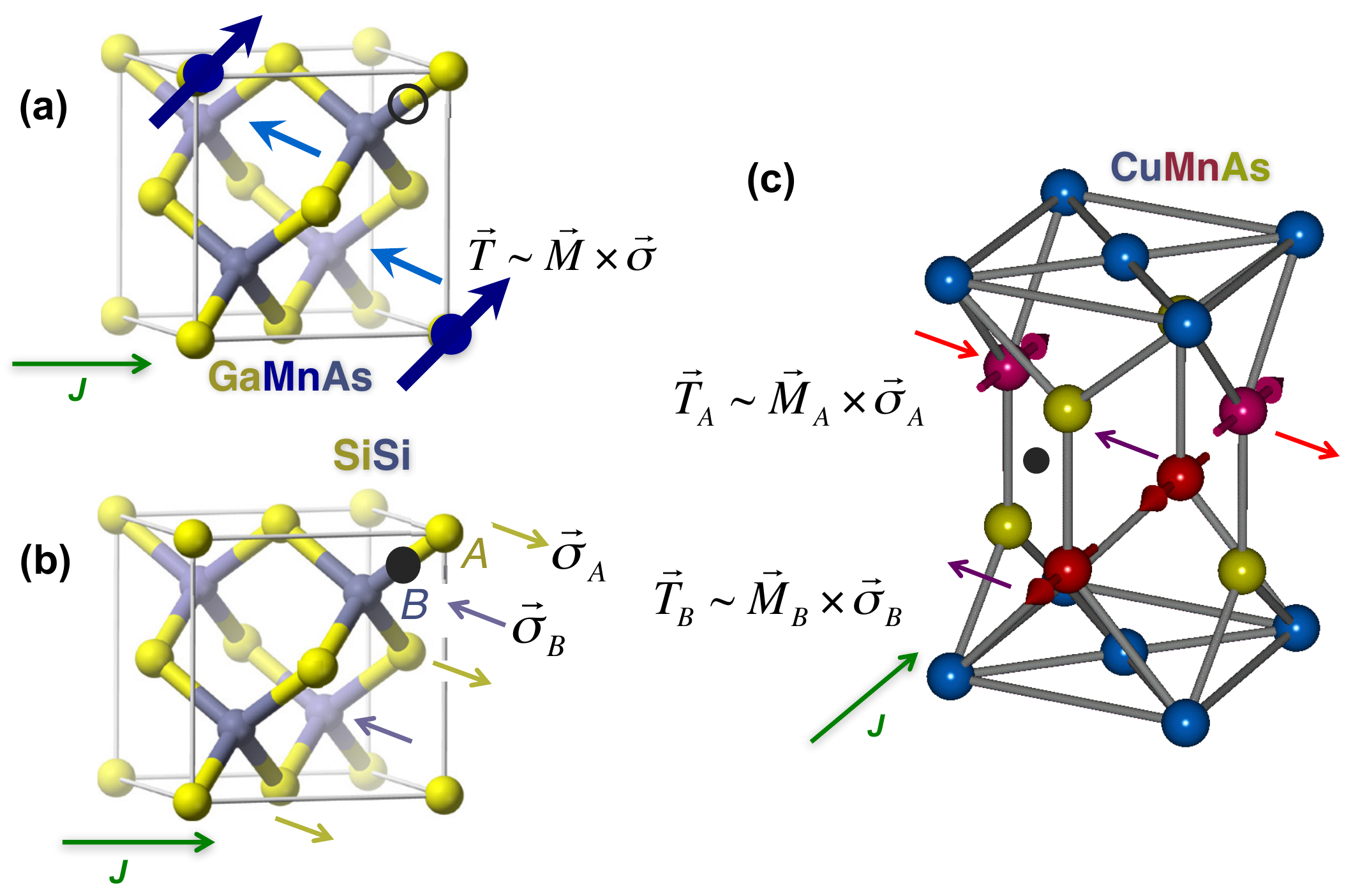}
	\caption[fig_1]{An illustration of iSGE induced spin-orbit torques in ferromagnets and antiferromagnets \cite{Ciccarelli2016,Jungwirth2016}. (a) Ferromagnetic iSGE torque is generated in crystals whose unit cell has globally broken inversion symmetry, like in (Ga,Mn)As. A charge current $J$ generates a uniform spin polarization $\boldsymbol{\sigma}$ which exerts a field-like torque $\mathbf{T}\propto\mathbf{M}\times\boldsymbol{\sigma}$ on the ferromagnetic moments. (b) An iSGE induced staggered spin polarization at locally non-centrosymmetric inversion-partner lattice sites A and B. Spin polarizations $\boldsymbol{\sigma}_\mathrm{A}$ and $\boldsymbol{\sigma}_\mathrm{B}$ have opposite sign at sites A and B. (c) An efficient, field-like N\'eel spin-orbit torque $\mathbf{T}_\mathrm{A,B}$  is generated in the antiferrmagnet when the antiferromagnetic spin-sublattices coincide with the inversion-partner crystal-lattice sites.}
	\vspace{-0.2 cm}
	\label{fig_1}
\end{figure}

When the SHE or iSGE spin polarization is strongly exchange coupled to spins forming an equilibrium magnetically-ordered state, a large internal effective field acts on the magnetic moments. The resulting relativistic spin-orbit torque can be used the electrically excite or even switch the magnet.  For example, in a non-magnetic/ferromagnetic bilayer, an in-plane electrical current driven through the non-magnetic layer generates the SHE spin-accumulation  at the interface with the ferromagnet where it exerts the spin-orbit torque on the ferromagnetic moments \cite{Manchon2008,Miron2010,Miron2011b, Liu2012c,Garello2013}. Similarly, the iSGE can induce an internal effective field and the resulting spin-orbit torque in a bulk ferromagnet \cite{Bernevig2005c,Chernyshov2009,Kurebayashi2014,Ciccarelli2016}. This is illustrated in Fig.~\ref{fig_1}(a) on a non-centrosymmetric lattice of the p-type (Ga,Mn)As ferromagnetic semiconductor. Here the electrical current produces the non-equilibruim iSGE spin-polarization in the As dominated valence band (thin arrows in Fig.~\ref{fig_1}(a)) which is exchange coupled to the ferromagnetic moments on Mn (thick arrows in Fig.~\ref{fig_1}(a)). When the non-equilibrium and the ferromagnetic spin-polarizations are misaligned, the internal iSGE field generates the torque on the magnetization. 

Antiferromagnets  have alternating magnetic moments in the ground state. The current-induced non-equilibrium spin polarization and the corresponding internal field that would efficiently couple to the antiferromagnetic order has to have a commensurate staggered order.  While the uniform field required in ferromagnets can be produced also by coils,  the generation of staggered fields in antiferromagnets is based purely on spin-orbitronics \cite{Zelezny2014}.   

To illustrate the formation of a staggered non-equilibrium spin polarization via the iSGE we take an example of a diamond lattice of, e.g., Si shown in Fig.~\ref{fig_1} (b) \cite{Zelezny2014,Zhang2014,Ciccarelli2016,Jungwirth2016}. The unite cell has two non-equivalent sites, A and B. Unlike in the zinc-blende lattice of GaAs (Fig.~\ref{fig_1} (a)), the two sites in the diamond lattice are occupied by the same chemical element. As a result, the unit cell has an inversion center between the sites A and B (black dot in Fig.~\ref{fig_1} (b)) and these two locally non-centrosymmetric sites are inversion partners. In such a globally centrosymmetric lattice, electrical current induces a zero net spin polarization when integrated over the whole unit cell. However, the current can generate non-zero local spin polarizations with equal magnitudes and opposite signs on the inversion-partner sites.  When this staggered iSGE spin polarization is exchange coupled to magnetic moments it results in a staggered internal effective field acting on the moments. The field is efficient in exciting magnetic moments when it is commensurate with the magnetic order, i.e., when the inversion partner crystal-lattice sites coincide with the opposite spin-sublattices of the antiferromagnet. This is the case of, e.g., CuMnAs shown in Fig.~\ref{fig_1}(c). In analogy with the established antiferromagnetic terminology where the staggered equilibrium magnetization is described as the N\'eel state, the non-equilibrium staggered field and the corresponding torque is called the N\'eel spin-orbit torque \cite{Zelezny2014}.

Electrical switching by the  N\'eel spin-orbit torque was recently demonstrated in  a CuMnAs memory chip at ambient conditions \cite{Wadley2016} and in a common microelectronic circuitry allowing to control the chip from a computer via a USB port \cite{Schuler2016}. The electrical readout was done by the AMR which is an even function of the magnetic moment and is, therefore, allowed by symmetry in antiferromagnets as well as in ferromagnets \cite{Daughton1992,Shick2010,Park2011b,Marti2014,Kriegner2016}. The antiferromagnetic memory bit cells show a multi-level switching characteristics associated with spin-orbit-torque controlled multiple-stable domain reconfigurations \cite{Schuler2016,Grzybowski2016}. This  allows in principle for storing more than one bit per cell and integrate memory and logic within the bit cell. For example, it was demonstrated that  a simple cross-geometry, micron-size bit cell alone can act as a neuron-like pulse-counter able to record thousands of pulses of lengths spanning eight orders of magnitude from $\sim 10$~ms down to a $\sim 100$~ps range \cite{Schuler2016}. The insensitivity of the writing, readout, and memory functionality of the CuMnAs chips was tested  in external magnetic fields up to 12~T \cite{Wadley2016}. It was also verified that these antiferromagnetic bit cells generate no stray fringing fields which is favorable for potential high-density-integration devices \cite{Wadley2016}. The bit cells were realized in CuMnAs films deposited at low temperature (200-300$^\circ$C) on Si and III-V substrates which opens the prospect of their utility in microelectronics and opto-electronics \cite{Schuler2016}.

\section{Antiferromagnetic order assisting ferromagnetic spintronics and vice versa}
\label{AF/FM}
Apart from the emerging utility of antiferromagnetic materials as active spintronic elements, antiferromagnetic order has traditionally played an important  supporting role in ferromagnetic spintronics. The discovery of the GMR (Nobel Prize in 2007) was made in the so-called synthetic antiferromagnets (SAFs) \cite{Baibich1988,Binasch1989}. They comprise alternating ferromagnetic and non-magnetic films in which carriers in the non-magnetic layers mediate an antiferromagnetic coupling among the ferromagnetic layers at zero external magnetic field. At an applied saturating field, the moments of the ferromagnetic films are switched into the parallel configuration, resulting in the GMR. 

A more robust and better controlled realization of the switching between antiparallel and parallel configurations is realized in GMR (TMR) stacks where one ferromagnet acts as a reference with a fixed magnetic moment orientation and  a second, free ferromagnet is the recording element. Crystal antiferromagnets are often used  in these structures to fix the magnetic moment in the reference ferromagnet by means of the inter-layer exchange bias \cite{Chappert2007}.  Remarkably, SAFs still play an important role in TMR bit cells of MRAMs. The reference part of the TMR stack is typically  made of a SAF in order to suppress the effect of stray fringing fields on the recording part of the stack and on neighboring cells \cite{Parkin2003}. SAFs are also used in the recording part of the commercial toggle-MRAM cells where they allow to realize a reliable spin-flop-like switching \cite{Engel2005a}.

\begin{figure}[h]
	\centering
	\includegraphics[width=1.00\columnwidth]{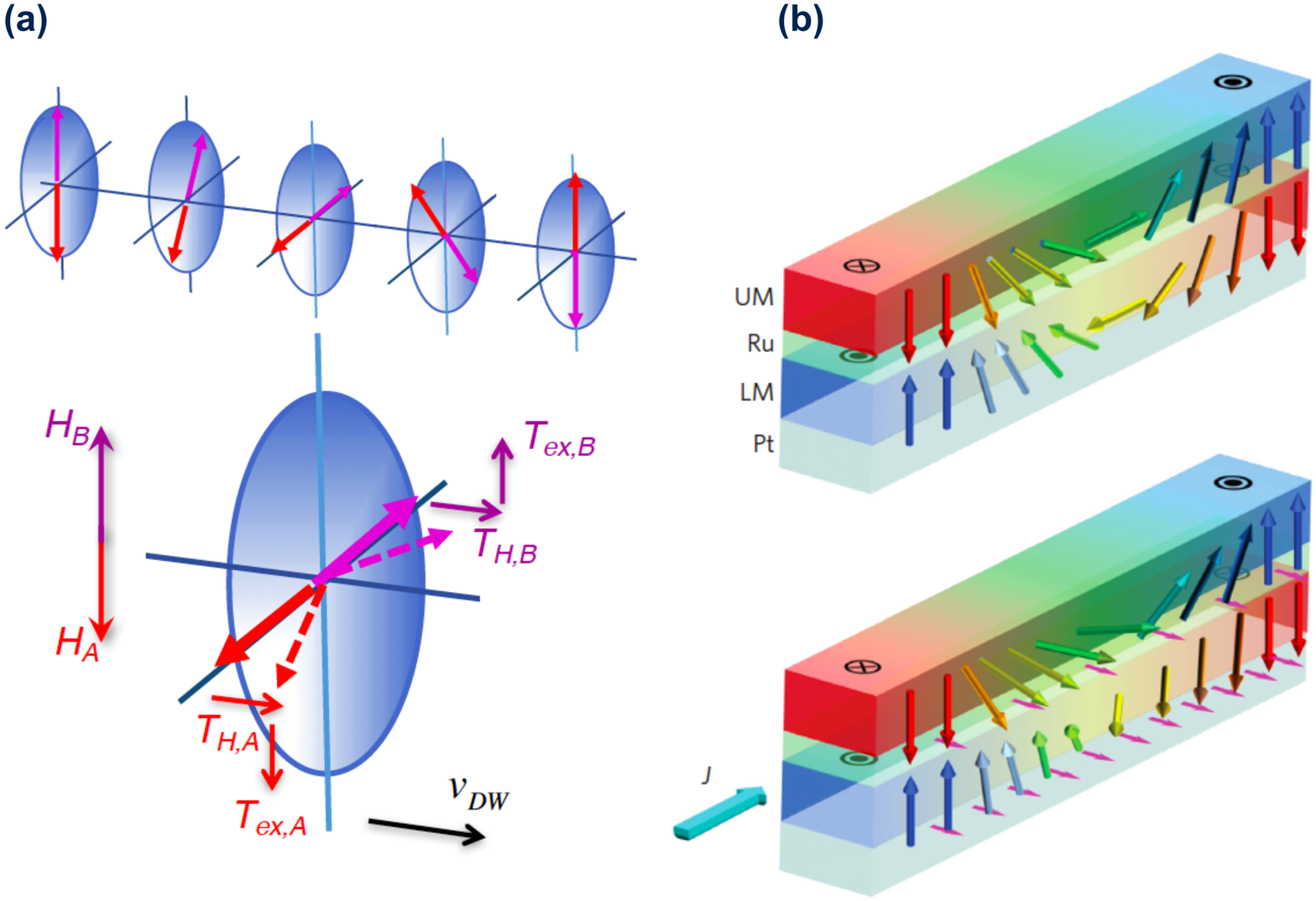}
	\caption[fig_2]{An illustration of the domain wall motion in a crystal antiferromagnet (a) \cite{Gomonay2016} and a SAF (b) \cite{Yang2015a}. Sublattice torques generated by a staggered field, $\mathbf{T}_\mathrm{H,A}$ and $\mathbf{T}_\mathrm{H,B}$, are parallel to each other and tend to tilt the sublattice magnetizations away from the domain wall plane. The resulting canting of the antiferromagnetic order is limited, however, by the strong inter-sublattice exchange coupling which delays in SAFs and eliminates in crystal antiferromagnets the Walker breakdown. Antiparallel torques $\mathbf{T}_\mathrm{ex,A}$ and $\mathbf{T}_\mathrm{ex,B}$ of the exchange nature then drive the domain wall motion.}
	\vspace{-0.2 cm}
	\label{fig_2}
\end{figure}

Antiferromagentic coupling in both crystal antferromagnets and SAFs is also favorable for achieving high domain wall velocities, as illustrated in Fig.~\ref{fig_2}.  For comparison, the domain wall velocity in ferromagnets is limited by the Walker breakdown \cite{Schryer1974}. When moving the ferromagnetic domain wall by a magnetic field or current, magnetization inside the wall is tilted away from the domain wall plane. At a critical Walker field, the static domain wall structure is broken and the resulting precession of spins  inside the domain wall hinders its motion. 

In crystal antiferromagnets,  the field that drives the domain wall motion tends to tilt the sublattice magnetizations towards the parallel configuration, as illustrated in Fig.~\ref{fig_2}(a) \cite{Tveten2014,Gomonay2016,Selzer2016,Shiino2016}. The resulting canting of the antiferromagnetic order in the perturbed domain wall is limited, however, by the strong inter-sublattice exchange coupling. The spin-flip field is required for the full ferromagnetic alignment of the spin-sublattices which in crystal antiferromagnets is excessively large, making the Walker breakdown unreachable. The domain wall velocity is then limited by the magnon velocity and can be as large as $\sim100$~nm/ps \cite{Tveten2014,Gomonay2016,Selzer2016,Shiino2016}. This means that the switching by the domain wall propagation of a $\sim100$~nm bit cell can occur at a picosecond scale which is two orders of magnitude faster than in ferromagnets.

Relatively high domain wall velocity can be also achieved in SAFs. In analogy to crystal antiferromagnets, the antiferromagnetic exchange coupling between the ferromagnetic layers  in the SAF acts against the canting-like domain wall deformation (see Fig.~\ref{fig_2}(b)) \cite{Yang2015a}. The exchange protection of the domain wall stability is, however, weaker in SAFs than in crystal antiferromagnets because the interlayer exchange coupling can only reach values that are two orders of magnitude smaller than the inter-sublattice exchange in crystal antiferromagnets \cite{Shiino2016}. Still, the Walker breakdown can be significantly delayed and the domain wall velocity limit increased from $\sim0.1$~nm/ps in bare ferromagnets to $\sim1$~nm/ps in SAFs \cite{Yang2015a}.  

\begin{figure}[h]
	\centering
	\includegraphics[width=1.00\columnwidth]{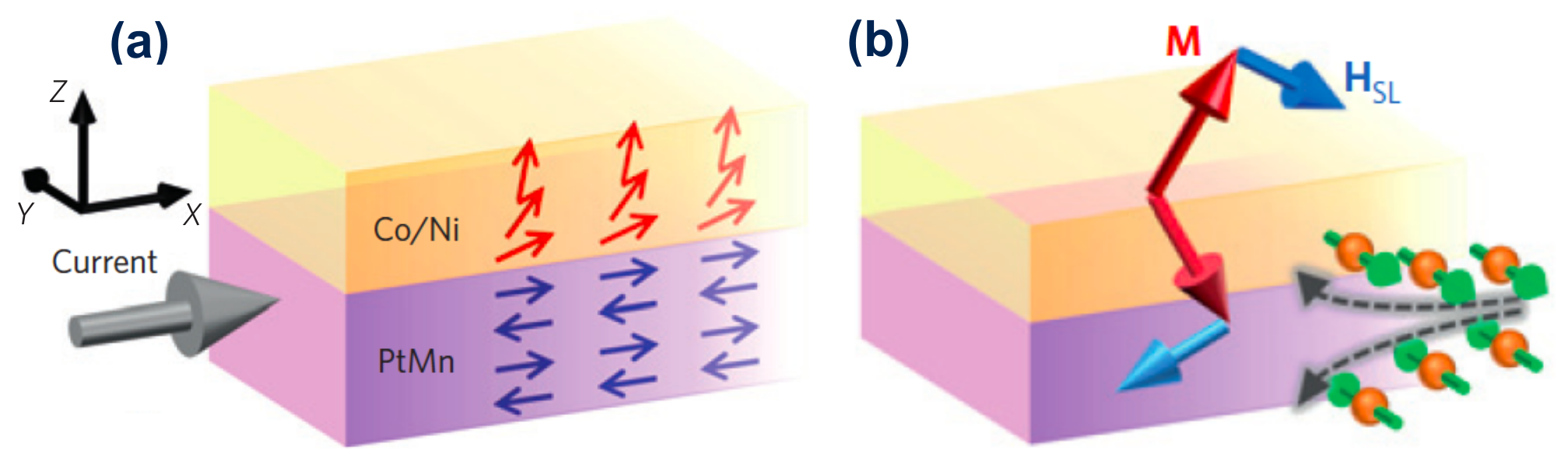}
	\caption[fig_3]{Deterministic spin-orbit torque switching of a ferromagent by an exchange-coupled antiferromagnet \cite{Fukami2016}. (a) Experimental geometry with the tilted equilibrium magnetization orientation in the CoNi ferromagnet resulting from a combination of the out-of-plane ($\hat{\mathbf{z}}$-axis) magnetoctystalline anisotropy and the in-plane ($\hat{\mathbf{x}}$-axis) unidirectional exchange-bias field. The in-plane electrical current generating the spin-orbit torque is driven along the exchange-bias field direction. (b) The current-induced SHE interfacial spin-polarization is along the $\hat{\mathbf{y}}$-axis (green arrows) and switching between up and down magnetized states is driven by the antidamping (Slonczewski) component of the corresponding spin-orbit torque, $\mathbf{T}\sim\mathbf{M}\times\mathbf{H}_{\rm SL}$, where $\mathbf{H}_{\rm SL}\sim\mathbf{M}\times\hat{\mathbf{y}}$. In the figure, the up-magnetized state is unstable while the down-magntized state is stable under the applied current. Upon reversing the current direction, the stable and unstable magnetization directions switch places since the SHE spin-polarization flips sign while the exchange-bias field  stays the same. Without the exchange-bias field, the spin-orbit torque switching is not deterministic.}
	\vspace{-0.2 cm}
	\label{fig_3}
\end{figure}

Antiferromagnets play also an important supporting role in current developments of ferromagnetic spin-orbit torque devices. Similar to strongly spin-orbit coupled paramagnets, antiferromagnets can host a large SHE \cite{Mendes2014,Zhang2014e}. For example, IrMn or PtMn antiferromagnets interfaced with a thin-film  transition-metal ferromagnet can generate spin-orbit torques in the ferromagnet with exceptionally high magnitudes \cite{Tshitoyan2015, Reichlova2015,Zhang2015f,Fukami2016,Zhang2016f,Oh2016,Borders2017}. 

Incidentally, these metallic antiferromagnets are extensively used in GMR/TMR stacks for spintronic applications where they pin the magnetization orientation  of the reference ferromagnet by the strong unidirectional exchange-bias field \cite{Chappert2007}, as mentioned at the beginning of this section.  By tuning the interfacial exchange coupling such that the exchange-bias field is still sizable but not strong enough to freeze the moments in the adjacent ferromagnet, it can be also utilized  as a  complementary tool that assists the spin-orbit torque switching in the recording part of the GMT/TMR stack.  This is illustrated in Fig.~\ref{fig_3} where the spin-orbit torque switching of the ferromagnet by the SHE from the antiferromagnet becomes deterministic due to the unidirectional exchange-bias field \cite{Fukami2016,Oh2016}.

Another remarkable feature of these ferromagnet/antiferromagnet spin-orbit torque devices is that they show a neuron-like multi-level swicthing characteristics \cite{Fukami2016,Borders2017}. In Sec.~\ref{NSOT}, we have already mentioned multi-level memory bit cells fabricated from a single-layer antiferromagnet CuMnAs \cite{Schuler2016}. Multi-level anistropic magnetoresistors have been also demonstrated in thin films of an antiferromagnetic semiconductor MnTe \cite{Kriegner2016}. In the above ferromagnet/antiferromagnet spin-orbit torque devices \cite{Fukami2016,Borders2017}, the recording medium is the ferromagnet. However, when the antiferromagnet is replaced in the stack with a paramagnetic SHE film, the spin-orbit-torque switched ferromagnet becomes bistable. This indicates that the multi-domain structure of the antiferromagnet is imprinted on the exchange-coupled ferromagnet and turns it into a multi-level medium. 
Antiferromagnets, whether alone or in combination with ferromagnets, thus open a prospect for developing neuron-like cells integrating memory and logic  and for realizing spintronics-based artificial intelligence \cite{Schuler2016,Borders2017}.  

Finally we recall that the ferromagnetic order  has also been used to assist the switching process in spintronic devices where the recording medium is an antiferromagnet \cite{Park2011b,Wang2012a,Petti2013,Marti2014,Fina2014}. An example is a tunneling device with a IrMn/NiFe recording electrode arranged in such a way that the antiferromagnetic IrMn film is adjacent to the tunnel barrier  \cite{Park2011b,Wang2012a,Petti2013}. In this configuration, the orientation of the antiferromagnetic moments determines the readout tunneling magnetoristance signal. The ferromagnetic film only serves a supporting role in the switching. Its moments are reoriented by an applied magnetic field while the antiferomagnetic moments respond indirectly to the magnetic field via the exchange spring effect at the ferromagnet/antiferromagnet interface. The observed antiferromagnetic tunneling AMR signals in these devices can exceed 100\% \cite{Park2011b} which is comparable to TMRs in ferromagnetic tunnel junctions and exceeds the typical AMR signals in ohmic device by two orders of magnitude \cite{Wadley2016,Grzybowski2016,Kriegner2016}. Large magnetoresistive readout signals are important for the size and speed scalability of the bit cells in highly integrated random access memories \cite{Daughton1992,Chappert2007}. 
   
\section{Ultrafast and topological antiferromagnetic spintronics}
\label{ultratopo}
\begin{figure}[h]
	\centering
	\includegraphics[width=1.00\columnwidth]{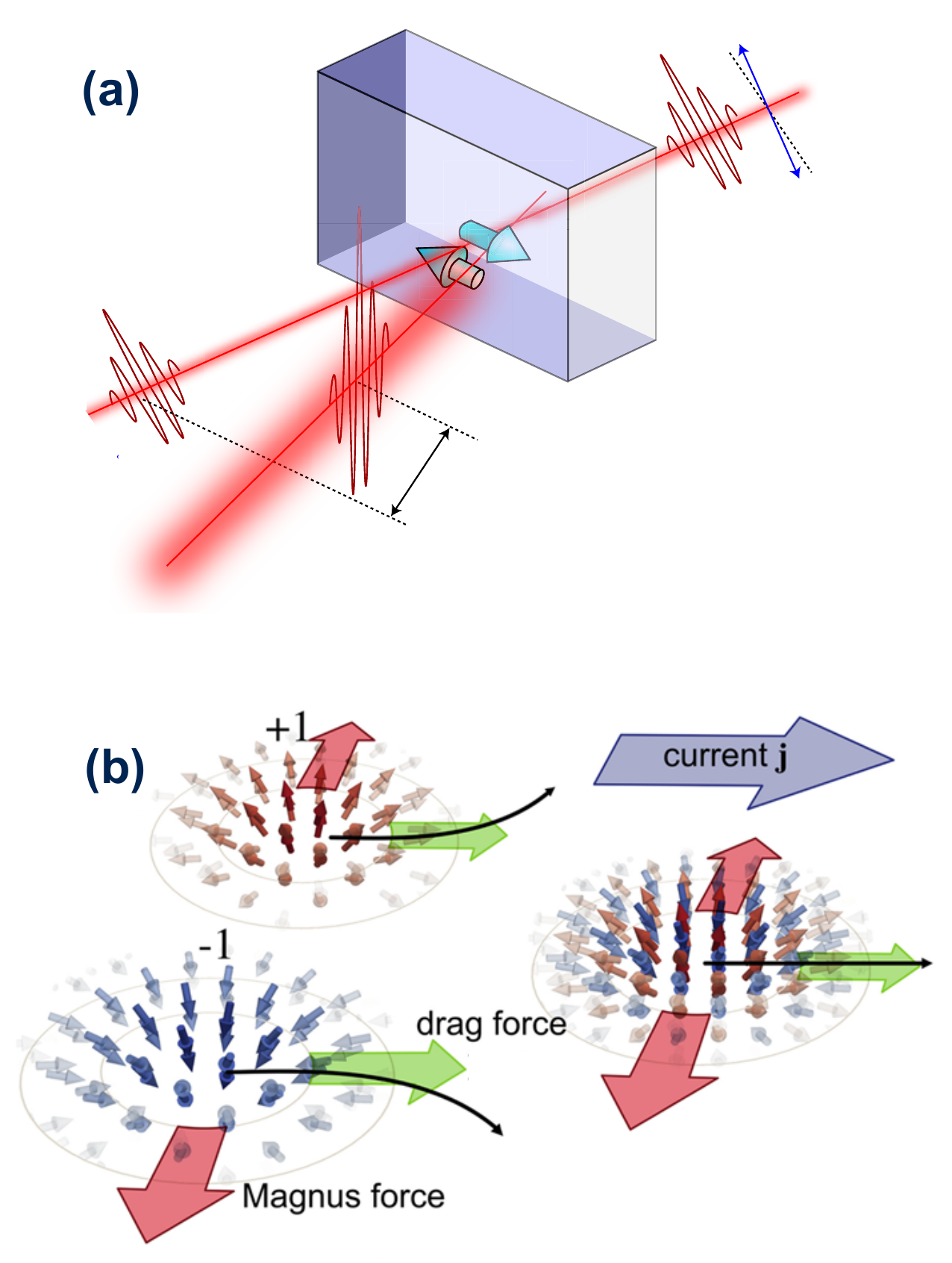}
	\caption[fig_4]{(a) Schematics of pump-probe laser set-up for optical excitation and detection of antiferromagnetic moments \cite{Saidl2017}. A more intense pump pulse is followed by a weaker, time-delayed probe pulse. (b) The antiferromagnetic skyrmion is composed of two topological objects with opposite
topological charge; hence, the Magnus force acts in opposite directions. The strong coupling between the antiferromagnetic spin-sublattices leads to a
perfect cancellation of the two opposing forces, and so, the  antiferromagnetic skyrmion has no transverse motion \cite{Barker2016}.}
	\vspace{-0.2 cm}
	\label{fig_4}
\end{figure}

The THz antiferromagnetic resonance \cite{Gomonay2014} has been directly observed in time-resolved pump-probe laser experiments, whose schematics is shown in Fig.~\ref{fig_4}(a), in various insulating materials. These include the prototypical antiferromagnet NiO \cite{Kampfrath2010}, a chiral three-sublattice complex oxide \cite{Satoh2014}, or a collinear fluoride antiferromagnet \cite{Bossini2015}.  In insulating antiferromagnets, intense laser pump pulses have been also shown to trigger ultrafast, (sub)picosecond transient switching of the N\'eel vector due to the light-induced easy-axis reorientation \cite{Duong2004,Kimel2004}. A reversible switching has been recently demonstrated in a helical antiferromagnet using a two-color-pump laser set-up \cite{Manz2016}.

Alternatively, theoretical predictions indicate \cite{Zelezny2014,Roy2016} that extending the N\'eel spin-orbit torque concept from the presently achieved $\sim100$~ps limit of electrically generated pulses \cite{Schuler2016} to optically controlled picosecond pulses can provide a straightforward route to the ultra-fast switching of antiferromagnets.  Optical readout of the N\'eel vector orientation in CuMnAs has been already demonstrated in the pump-probe set-up (Fig.~\ref{fig_4}(a)) using the magnetic linear dichroism (an ac counterpart of the AMR) \cite{Saidl2017}. It opens the prospect of bridging the fields of electronic and optical spintronics in a metallic antiferromagnet. Apart from ultra-fast memories, antiferromagnets are also natural candidates for realizing THz spin-torque oscillators \cite{Gomonay2008, Cheng2015a}.

Research in antiferromagnetic spintronics is not limited to uniform systems. Antiferromagnetic nanostructures can be manipulated and detected with atomic resolution which was demonstrated in spin-polarized scanning tunneling microscopy experiments  \cite{Wieser2011,Loth2012}. Antiferromagnetic textures, like the skyrmions shown in Fig.~\ref{fig_4}(b), have intriguing topological properties \cite{Barker2016,Zhang2015b}. An antiferromagnetic skyrmion is a compound topological object with a similar but of opposite sign spin texture on each sublattice which results in a complete cancellation of the Magnus force (Fig.~\ref{fig_4}(b)). Unlike in ferromagnets, skyrmions in antiferromagnets should, therefore, move in straight lines along the driving current. Another example of intriguing phenomena arising in non-collinear antiferromagnetic structures is the topological or anomalous Hall effect \cite{Surgers2014,Chen2014,Nakatsuji2015,Kiyohara2015,Nayak2016}.
 
Finally we turn to topological phenomena in the k-space of the band-structure of collinear antiferromagnets \cite{Smejkal2016}. Dirac electrons and other quasiparticles mimicking different flavors of fermions from relativistic particle physics, are now recognized as an intriguing new platform for exploring topologically protected phases in condensed matter \cite{Qi2011}.  While fascinating theoretically, practical means for controlling these phases in devices have remained elusive. Spintronics could be a key here, however, ferromagnets always break the $\cal{P}$(space-inversion)-$\cal{T}$(time-reversal)-symmetry required for the formation of topological Dirac fermions. While also breaking the $\cal{T}$-symmetry, antiferromagnets can preserve the combined $\cal{PT}$-symmetry when the opposite spin-sublattices occupy space-inversion partner lattice sites \cite{Tang2016,Smejkal2016}. Serendipitously, this condition coincides with the symmetry requirement for the efficient N\'eel spin-orbit torque in the antiferromagnet driven by the staggered, current-induced effective field that is commensurate with the staggered antiferromagnetic order (Fig.~\ref{fig_1}(c)) \cite{Zelezny2014,Smejkal2016}. This suggests a new concept in which the control of topological relativistic quasiparticles is mediated by reorienting the magnetic order parameter and where the efficient means for the magnetic moment reorientation is provided by antiferromagnetic spintronics.  New phenomena may arise from this concept, including the topological metal-insulator transition and AMR \cite{Smejkal2016}. This is an illustration of how antiferromagnetic spintronics can built bridges between seemingly incompatible fields of condensed matter physics and circumvent seemingly fundamental limitations for their utility in future technology.

\section{Outlook}
\label{outlook}
The concepts discussed above illustrate that after nearly a century of obscurity, antiferromagnets are turning into a new paradigm for intertwined science and technology. In science, antiferromagnets may become a unifying platform for realizing synergies among three prominent fields of contemporary condensed matter physics, namely spintronics (Nobel Prize 2007), Dirac quasiparticles (Nobel Prize 2010), and topological phases (Nobel Prize 2016). 

While the course of scientific discoveries is intrinsically unpredictable, information technologies are now entering a comparably unpredictable era which makes them equally exciting. Previous decades strived to build compact computer boxes equipped with highly-integrated devices for ever increasing data processing and storage capacity, assuming unlimited power resources. This relatively simple landscape ceased to exist with the end of the Moore's Law and with the invasion of mobile, internet-of-things, and cloud technologies \cite{Waldrop2016}. As of 2016, the Moore's Law driven International Technology Roadmap for Semiconductors is officially at an end \cite{Waldrop2016}. It is being replaced with the new  International Roadmap for Devices and Systems \cite{Waldrop2016} in order to tackle the scaling problem that is further magnified by the huge increase in the complexity of information technologies.  

In the emerging internet-of-things era, the "computer" is out of the box  with its billions of bit-cells dispersed in the streets, buildings, and vehicles, each playing a specialized role with a minimum energy consumption and a robust and secure performance. It seems unlikely that a single technological approach, and antiferromagnetic spintronics is no exception here,  can cope with such a diverse demand. However, the range of unique characteristics including the non-volatility, radiation and magnetic-field hardness, no fringing stray fields, THz spin-dynamics, or the neuron-like memory-logic functionality make antiferromagnetic spintronics one of the new concepts prone to make a mark in the "More than Moore" technologies \cite{xavi}.

The authors acknowledge the Alexander von Humboldt Foundation, the ERC Synergy Grant SC2 (No. 610115), the Transregional Collaborative Research Center (SFB/TRR) 173 SPIN+X, the Grant Agency of the Czech Republic Grant No. 14-37427G, and  the Ministry of Education of the Czech Republic Grant No. LM2015087.
%

\end{document}